\documentclass{PoS}

\usepackage{amsmath,amsfonts,amssymb}
\newcommand{\ptmiss}{p_T\!\!\!\!\!\!\!\!\not\,\,\,\,\,\,\,}
\newcommand{\SM}{\Sigma_\text{M}}
\newcommand{\SD}{\Sigma_\text{D}}

\title{New neutrino interactions at large colliders}

\ShortTitle{New neutrino interactions at large colliders}

\author{\speaker{Francisco del \'Aguila}, 
Juan Antonio Aguilar-Saavedra 
\thanks{This work has been partially supported by MICINN (FPA2006-05294), 
Junta de Andaluc{\'\i}a (FQM 101, FQM 03048) and the European Union 
(MRTN-CT-2006-035505).} \\
Depto. de Fsica Te\'orica y del Cosmos and CAFPE, U. de Granada, E-18071 Granada, Spain\\
E-mail: \email{faguila@ugr.es},
\email{jaas@ugr.es}}

\author{Jorge de Blas\\
Department of Physics, University of Notre Dame,
Notre Dame, Indiana 46556, USA \\
        E-mail: \email{jdeblasm@nd.edu}}

\abstract{We update present bounds on the helicity of the neutrinos produced 
in muon decay, including $e^+e^- \rightarrow \overline{f}f$ LEP 2 data. 
These significantly reduce the limits derived from all the other electroweak 
precision data. In Standard Model extensions designed to maximize the RH neutrino 
production in such a decay the neutrino deficit eventually observable in a near 
detector at a neutrino factory can be of the order of 5 \%.  
Motivated by the current LHC run at 7 TeV, we also update previous work providing 
discovery limits on see-saw mediators at this centre of mass energy. 
Lepton triplets with 200 GeV could be discovered with luminosities of 1 - 1.5 fb$^{-1}$. 
Scalar triplets of the same mass might be seen with 0.75 - 3 fb$^{-1}$. 
What makes their search also attractive in the first LHC analyses.}

\FullConference{35th International Conference of High Energy Physics - ICHEP2010,\\
		July 22-28, 2010\\
		Paris France}

\begin{document}

\section{Introduction}

Neutrino masses and neutrino interactions manifest at two very different energy 
scales. Thus, while neutrino masses are in the eV range or below, weak 
interactions occur near the TeV, at a scale 10$^{12}$ times larger. 
As a consequence, neutrino masses have only showed up at neutrino oscillation 
experiments where a long baseline enhances their effect 
\cite{Nakamura:2010zzi}, 
but they can be completely neglected in experiments where the relevant energies 
are of the electroweak order, $v \sim 246$ GeV, as in large collider experiments. 
This is the whole story within the minimal Standard Model (SM) extension with tiny 
Dirac neutrino masses (Yukawa couplings), or with small Majorana neutrino masses 
from a highly suppressed dimension five Weinberg operator 
$({\cal O}_5)_{ij}= \overline{(l_L^i)^c}\tilde \phi ^* \tilde \phi^\dagger l_L^j$ 
(see-saw mechanism) 
\cite{Mohapatra:1998rq}. 
However, this situation can be totally reversed if new interactions involving 
light left (LH) and/or right-handed (RH) neutrinos exist near the electroweak 
scale \cite{Nath:2010zj}. 

In this short communication we revise two interesting cases. 
The possibility that there are light Dirac or Majorana RH neutrinos with 
appropriate new weak interactions which may be observable at a large collider 
and, in particular, at a neutrino factory 
\cite{delAguila:2009vv}, 
or on the contrary, that there are no light RH neutrinos but the see-saw 
mediators of the light Majorana masses for LH neutrinos live near the TeV 
at the LHC reach 
\cite{delAguila:2008cj}. 

\subsection{Observable RH neutrino production}

\begin{figure}
\includegraphics[width=.5\textwidth]{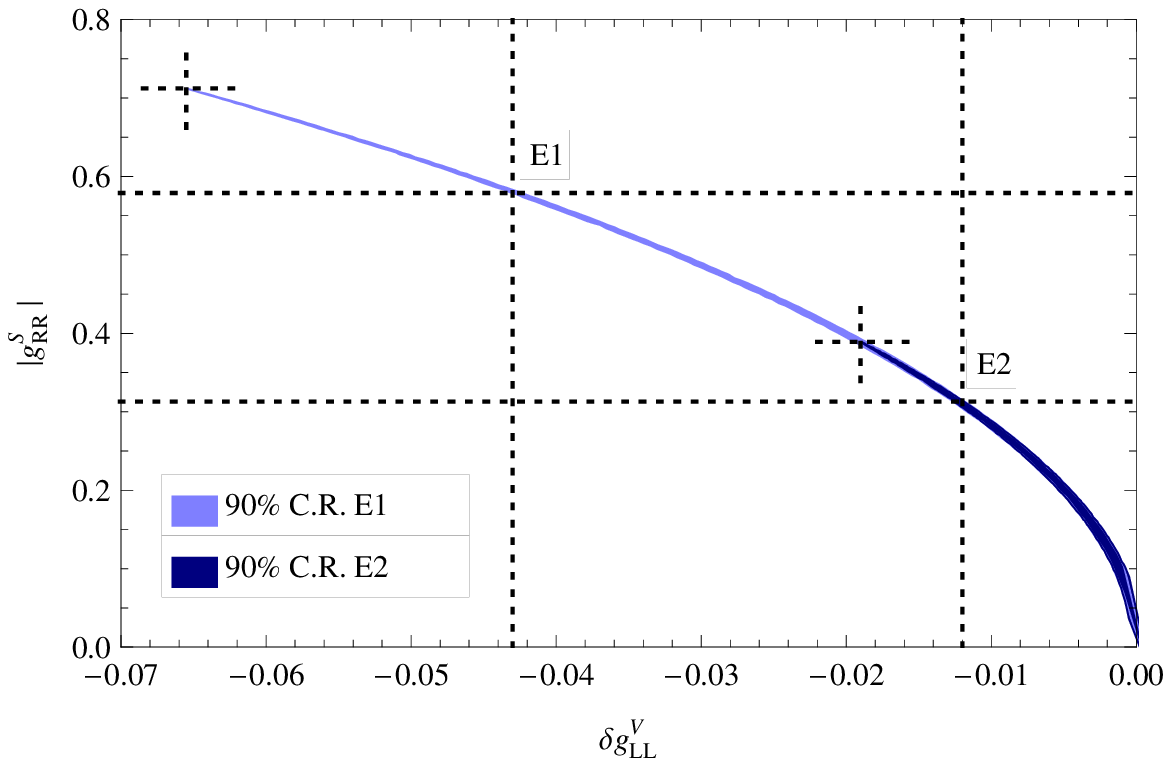}
\includegraphics[width=.5\textwidth]{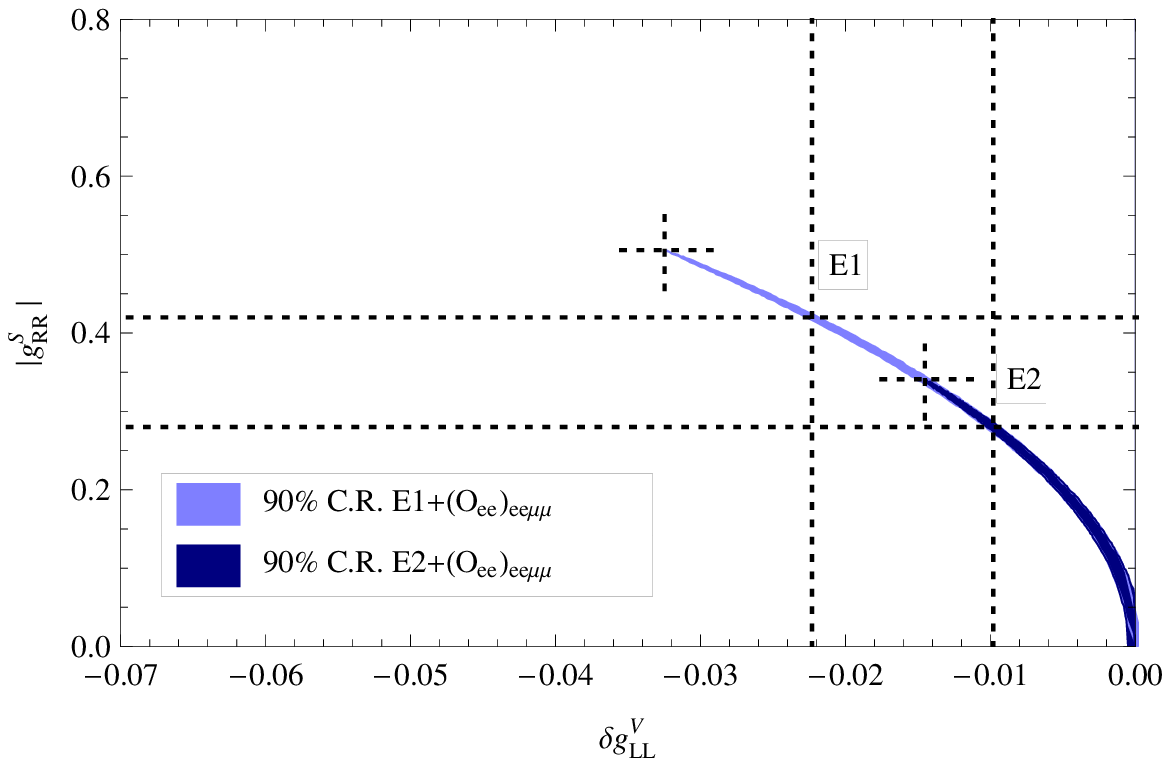}
\caption{(Left) 90 \% C.L. bounds from a global fit using EWPD (without LEP 2 data)
to appropriate extra dimension six operators involving light LH (abscisse) and RH
(ordinate) neutrinos. E1 and E2 stand for the model independent and the specific
model addition described in [4], respectively. (Right) The same including also LEP 2
data in the fit and considering the extra four lepton operator $({\cal
O}_{ee})_{ee\mu\mu}$. The crosses define the extremes of the corresponding 90 \%
confidence regions for the global fit to the new parameters.}
\label{fig1}
\end{figure}

A fair question is how well do we know that light neutrinos are LH ? 
or to be more concrete, which helicity have the neutrinos produced in 
muon decay ? 
\cite{Langacker:1988cm}. 
From pion decay into muons we know that the muon charged interactions 
are mainly LH, and electroweak precision data (EWPD) require this to 
be so with a precision of few per ten thousand within the SM 
\cite{Nakamura:2010zzi}. 
This is reflected by the width of the curves in Fig. \ref{fig1}. 
However, if we allow for light RH neutrinos and new four-fermion 
interactions of the electroweak size up to a factor of a few, and 
possible model dependent cancellations, present experimental 
constraints leave room for light RH neutrinos observable at a 
neutrino factory, and new mediators observable at LHC 
\cite{delAguila:2009vv,delAguila:2010zh}. 
Indeed, although limits on new physics described by dimension six 
operators are typically of the order of 1-10 \% 
\cite{Biggio:2009nt}, if we allow for cancellations some deviations from 
the SM can be larger. In Fig. \ref{fig1} (left) we show two examples 
  (see \cite{delAguila:2009vv,deBlas2010} for definitions and details) where the 
large deviations correspond to the relatively weak bounds 
(straight lines and crosses) on $\delta g_{LL}^V$ and $g_{RR}^S$. 

It must be noticed, however, that although these SM deviations 
could be observable at a large $\nu$ factory as a deficit of the neutrinos 
detected by a near detector of at most 8.5 (4.5) \% through inverse muon decay 
(neutrino-nucleon) and an adequate, model independent dimension 
six operator addition (case E1 in Fig. \ref{fig1} (right)), such an addition is 
only designed to maximize this effect while preserving present 
experimental limits, including those from rare processes. 
When one also demands that such an addition results from a definite 
model, the possible deviations are reduced (for instance, to 4 (2) \% 
in case E2 in Fig. \ref{fig1} (right)). Finally, it must be stressed that 
these departures from the SM predictions are compatible with 
all EWPD, including in particular $e^+e^- \rightarrow \overline{f}f$ LEP 2 
data (right panel in Fig. \ref{fig1}). 
These appear to be quite restrictive, not only reducing the new 
parameter space as shown by comparing the left and right panels 
in Fig. \ref{fig1}, but requiring extra additions to maintain the agreement 
with experimental data while allowing for relatively large 
$\delta g_{LL}^V$ and $g_{RR}^S$ values, for example 
$({\cal O}_{ee})_{ee\mu\mu} = 
\frac{1}{2}(\overline{e_R}\gamma_\alpha e_R) (\overline{\mu_R}\gamma^\alpha \mu_R)$ 
in the two cases depicted in Fig. \ref{fig1} (right). 

\subsection{Observable see-saw mediators at LHC}

\begin{figure}
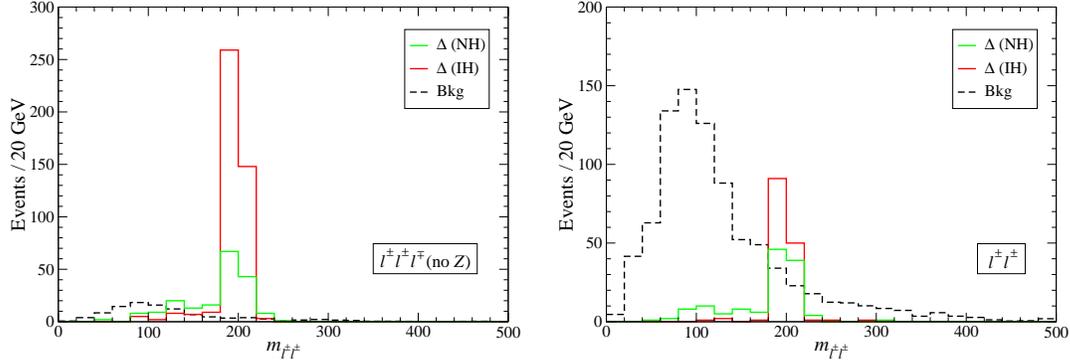

\quad \includegraphics[width=.45\textwidth]{mll-3Q1.eps} \quad
\includegraphics[width=.45\textwidth]{mll-2Q2.eps}
\caption{Reconstructed scalar triplet mass 
in the $\ell^\pm \ell^\pm \ell^\mp$ (no $Z$) (left) and  
in the $\ell^\pm \ell^\pm$ (right) samples 
for a center of mass energy of 7 TeV and an integrated luminosity of 30 fb$^{-1}$.}
\label{fig:mrec-3Q1-sc}
\end{figure}

Light neutrinos may be only LH and get their masses through the Weinberg operator  
$({\cal O}_5)_{ij}$ after spontaneous symmetry breaking. 
This dimension five operator results from the tree-level exchange of 
fermion singlets $N$ and/or triplets $\Sigma$, and/or scalar triplets $\Delta$. 
Which are known as type I, III and II see-saw mechanism, 
respectively 
\cite{Mohapatra:1998rq,Nath:2010zj,delAguila:2009bb}.  
In this case if these heavy mediators have masses near the TeV, 
being then eventually observable at LHC, 
the smallness of the light neutrino masses requires 
an extremely small $\sim 10^{-12}$ effective coupling $x_{ij}$ 
multiplying $({\cal O}_5)_{ij}$. What stands for very small mixing 
angles in the fermionic case, $|V_{\ell N, \ell \Sigma}| \sim 10^{-6}$, or a 
minuscule effective lepton number breaking parameter in the 
scalar one, $\sim 10^{-12}$. Such small numbers translate into 
negligible mediator production cross sections at large colliders 
if their production mechanism is proportional to them, as in the 
fermion singlet case in the absence of further interactions 
\footnote{Although the mixing angle $V$ is decoupled from the 
light neutrino masses, as in the case of quasi-Dirac heavy neutrinos, 
EWPD require 
$|V_{e N, \mu N, \tau N}| < 0.05, 0.03, 0.09$ at 95 \% C.L. 
\cite{deBlas2010,delAguila:2009bb,delAguila:2008pw}. 
What makes difficult to observe them at LHC 
\cite{delAguila:2008cj,delAguila:2008hw}.}. 
(Otherwise, its production rate can be of the electroweak size, 
as for instance in Left-Right models 
\cite{Nath:2010zj,delAguila:2009bb}.) 
In the triplet case their production is through electroweak 
interactions. In such a case they are observable at LHC for 
masses below the TeV.  
Here we present the corresponding results for a center of mass energy of 7 TeV. 
In Table \ref{tab:summ} we gather the luminosity required 
for a $5\sigma$ discovery of the see-saw triplet messengers in 
three different multi-lepton channels. 
As for a center of mass energy of 14 TeV 
\cite{delAguila:2008cj,delAguila:2008hw}, the four cases considered can be 
distinguished by comparing the number of events in the different samples. 
The pre-selection and selection criteria are also similar, with a good 
reconstruction of the heavy resonance masses, as shown in Fig. \ref{fig:mrec-3Q1-sc} 
for scalar triplets with a mass of 200 GeV and a luminosity of 30 fb$^{-1}$.
\begin{table}
\begin{center}
\begin{tabular}{cccc}
      & $\ell^\pm \ell^\pm \ell^\mp$ (no $Z$) & 
$\ell^\pm \ell^\pm$ (no $\ptmiss$) & $\ell^+ \ell^+ \ell^- \ell^-$ \\
$\Delta$ (NH) & 2.7  & 5.9 & 11.5 \\
$\Delta$ (IH) & 0.74 & 2.1 & 2.0  \\
$\SM$         & 2.0  & 1.4 & 3.6  \\
$\SD$         & 0.97 & --  & 1.0  \\
\end{tabular}
\end{center}
\caption{Luminosities (in fb$^{-1}$) required for $5\sigma$ discovery 
at LHC with a center of mas energy of 7 TeV 
for the models in the left column (scalar triplets $\Delta$ coupling 
to light neutrinos with a normal NH or an inverted IH mass hierarchy, 
and Majorana M or Dirac D fermion triplets $\Sigma$) in the final states 
indicated. A dash stands for an unobservable signal.}
\label{tab:summ}
\end{table}

\end{document}